\documentclass[oneside,english]{amsart}
\usepackage[T1]{fontenc}
\usepackage[latin9]{inputenc}
\usepackage{amsthm}
\usepackage{amstext}
\usepackage{amssymb}
\usepackage{graphicx}

\makeatletter
\numberwithin{equation}{section}
\numberwithin{figure}{section}
\theoremstyle{plain}
\newtheorem{thm}{\protect\theoremname}
  \theoremstyle{remark}
  \newtheorem{rem}[thm]{\protect\remarkname}
\newenvironment{lyxlist}[1]
{\begin{list}{}
{\settowidth{\labelwidth}{#1}
 \setlength{\leftmargin}{\labelwidth}
 \addtolength{\leftmargin}{\labelsep}
 }}
{\end{list}}

\makeatother

\usepackage{babel}
  \providecommand{\remarkname}{Remark}
\providecommand{\theoremname}{Theorem}

\begin{document}

\title{An analysis of Rick Lockyer's ``octonion variance sieve''}

\author{Jens K\"oplinger}

\address{105 E Avondale Dr, Greensboro NC, 27403}

\email{jens@prisage.com}

\urladdr{http://jenskoeplinger.com/P}

\date{23 March 2011 (v1), 5 December 2012 (v2)}
\begin{abstract}
In ``Octonion Algebra and its Connection to Physics'' \cite{LockyerOctonions-Feb2008}
an algorithm on octonions is brought forward for description of physical
law, the ``octonion variance sieve process''. This paper expresses
the used algorithm in symbolic form, and highlights the structure
between the ``function'', ``distance'', and ``algebraic invariant''
concepts therein. An alternative description in terms of derivation
algebras is shown.
\end{abstract}
\maketitle

\section{Introduction}

Maxwell electromagnetism has been expressed various times on octonionic
algebras (e.g.~\cite{Gogb2006OctonionElectrodyn,CandemirEtAlHyperbolicProcaMW,TolanEtAtReformEMOct,NurowskiSplitOctMaxw2009}),
and octonionic Dirac equations or spinors (e.g.~\cite{SchrayManogueOcts1994,Okubo1995IntoOctoNonassocBook,GunaydinGursey1974QuarksOcts,Dixon1994DivisionAlgs,DrayManogue1998DimRed,Dzhun2009Hidden,BaezHuerta2009DivisionAlgs1,Furey2010UnifThIdeal,KoeplDzhun2009NonassocQTcoquasigroup,DeLeo1996Progr1,DeLeo1996Progr2})
promise usefulness of octonions across all fundamental forces in physics.
A recent proposal, the ``octonion variance sieve'' in \cite{LockyerOctonions-Feb2008}
introduces a class of functions on the octonions that are invariant
under a new algorithm. A ``law of octonion algebraic invariance''
recovers the general electromagnetic action when assigning the electromagnetic
field to certain octonion functions, and modeling dynamic interaction
through octonionic differential operators.

The current paper restates this octonion variance sieve in a more
symbolic form and highlights structural similarities between concepts
used therein. Generalization to connection to classical electrodynamics
(as then treated in \cite{LockyerOctonions-Feb2008}) or a future
nonassociative quantum theory (as in \cite{KoeplDzhun2009NonassocQTcoquasigroup})
will not be handled here.

\section{Octonions}

The octonions $\mathbb{O}$ are the highest-dimensional normed division
algebra. They supply $\mathbb{R}^{8}$ with a multiplicative norm
$\left\Vert \cdot\right\Vert :\mathbb{R}^{8}\rightarrow\mathbb{R}_{0}^{+}$,
where for any $a,b\in\mathbb{O}$ there is $\left\Vert a\right\Vert \left\Vert b\right\Vert =\left\Vert ab\right\Vert $.
The multiplicative inverse for any number other than $0$ is unique,
$\forall a\in\mathbb{O}/\left\{ 0\right\} \exists b,\, ab=1$. Using
eight orthogonal vectors in $\mathbb{R}^{8}$ as octonion basis,
\begin{eqnarray}
b_{\mathbb{O}} & := & \left\{ 1,i_{1},\ldots,i_{7}\right\} ,
\end{eqnarray}
an octonion is described through real coefficients $a:=\left(a_{0},\ldots,a_{7}\right)$.
``Addition'' is the vector space addition, ``multiplication''
distributes over addition and is described by basis element relations
$1^{1}=1$, $i_{n}^{2}=-1$ ($n=1,\ldots,7$), and a set of seven
associative anticommutative ordered triplets $t_{\mathbb{O}\left[N\right]}$:
\begin{eqnarray}
i_{l}i_{m} & = & \epsilon_{lmn}i_{n}\qquad\textrm{for all }\left\{ l,m,n\right\} \in t_{\mathbb{O}\left[N\right]}.
\end{eqnarray}
These associative triplets then fully describe an octonion algebra,
e.g.:
\begin{eqnarray}
 &  & t_{\mathbb{O}\left[0\right]}:=\label{eq:defTO0}\\
 &  & \left\{ \left\{ 1,2,3\right\} ,\left\{ 7,6,1\right\} ,\left\{ 5,7,2\right\} ,\left\{ 6,5,3\right\} ,\left\{ 1,4,5\right\} ,\left\{ 2,4,6\right\} ,\left\{ 3,4,7\right\} \right\} .\nonumber 
\end{eqnarray}
Even permutations within a triplet do not change the algebra and are
choice of notation only. A total of 16 sets of triplets, $t_{\mathbb{O}\left[N\right]}$,
$N=0,\ldots,15$ now generates equivalent octonion multiplication
rules $\mathbb{O}\left[N\right]$, under the class that allows for
odd permutations within some of the triplets. The $\mathbb{O}\left[N\right]$
are called \emph{equivalent algebras} for short in this paper. Notation
will be abbreviated for the $t_{\mathbb{O}\left[N\right]}$ by using
the octonion index numbers from $t_{\mathbb{O}\left[0\right]}$ (equation
\ref{eq:defTO0}) as a reference, and then indicating whether permutations
within each triplet are of even ($+$) or odd ($-$) parity. For example:
\begin{eqnarray}
 &  & t_{\mathbb{O}\left[1\right]}:=\left\{ +,+,-,-,+,-,-\right\} \\
 & \equiv & \left\{ \left\{ 1,2,3\right\} ,\left\{ 7,6,1\right\} ,\left\{ 5,2,7\right\} ,\left\{ 6,3,5\right\} ,\left\{ 1,4,5\right\} ,\left\{ 2,6,4\right\} ,\left\{ 3,7,4\right\} \right\} .\nonumber 
\end{eqnarray}

\section{Automorphisms}

Four duality automorphisms, $\mathcal{T}_{0},\ldots,\mathcal{T}_{3}$,
act on the multiplication rules $\mathbb{O}\left[N\right]$ that make
up equivalent octonion algebras:
\begin{eqnarray}
\mathcal{T}_{n} & : & \left\{ \mathbb{O}\left[N\right]\right\} \rightarrow\left\{ \mathbb{O}\left[N\right]\right\} ,\qquad n\in\left\{ 0,1,2,3\right\} ,\\
\mathcal{T}_{n}\mathcal{T}_{n} & = & \left(\mathrm{id}\right),\nonumber \\
\left\{ \left(\mathrm{id}\right),\mathcal{T}_{n}\right\}  & \cong & \mathbb{Z}_{2}.\nonumber 
\end{eqnarray}
$\mathbb{Z}_{2}$ is the cyclic group with two elements. Acting on
the $t_{\mathbb{O}\left[N\right]}$, the $\mathcal{T}_{n}$ either
leave the parity of a permutation triplet unchanged, $\left(\mathrm{id}\right)$,
or swap it, $\left(\mathrm{sw}\right)$:
\begin{eqnarray}
\mathcal{T}_{0} & := & \left\{ \left(\mathrm{id}\right),\left(\mathrm{id}\right),\left(\mathrm{id}\right),\left(\mathrm{id}\right),\left(\mathrm{sw}\right),\left(\mathrm{sw}\right),\left(\mathrm{sw}\right)\right\} ,\label{eq:defAutomorphismsT}\\
\mathcal{T}_{1} & := & \left\{ \left(\mathrm{sw}\right),\left(\mathrm{sw}\right),\left(\mathrm{sw}\right),\left(\mathrm{sw}\right),\left(\mathrm{id}\right),\left(\mathrm{id}\right),\left(\mathrm{id}\right)\right\} ,\nonumber \\
\mathcal{T}_{2} & := & \left\{ \left(\mathrm{id}\right),\left(\mathrm{sw}\right),\left(\mathrm{id}\right),\left(\mathrm{sw}\right),\left(\mathrm{sw}\right),\left(\mathrm{id}\right),\left(\mathrm{sw}\right)\right\} ,\nonumber \\
\mathcal{T}_{3} & := & \left\{ \left(\mathrm{id}\right),\left(\mathrm{id}\right),\left(\mathrm{sw}\right),\left(\mathrm{sw}\right),\left(\mathrm{id}\right),\left(\mathrm{sw}\right),\left(\mathrm{sw}\right)\right\} .\nonumber 
\end{eqnarray}
Whereas $\mathcal{T}_{0}$ changes the parity of three triplets, the
$\left\{ \mathcal{T}_{1},\mathcal{T}_{2},\mathcal{T}_{3}\right\} $
each change the parity of four triplets. $\mathcal{T}_{0}$ transitions
between what is called ``left-'' and ``right-handed'' octonion
multiplication rules \cite{LockyerOctonions-Feb2008}, that are ``not
isomorphic'' in the sense that they cannot be transformed into one
another through transformation of the basis vectors in $\mathbb{R}^{8}$
alone. Instead, $\mathcal{T}_{0}$ is an algebra isomorphism that
transitions between opposite algebras of different chirality \cite{SchrayManogueOcts1994}.
The combined $\mathcal{T}_{0}\mathcal{T}_{1}$ inverts the sign of
all seven nonreal octonion basis elements and corresponds to complex
conjugation.

\begin{figure}
\centering{}\caption{\label{fig:T1T2T3Fano}All unique automorphisms from repeat application
of the $\left\{ \mathcal{T}_{1},\mathcal{T}_{2},\mathcal{T}_{3}\right\} $
can be graphed in the Fano plane (left), where the product of each
two automorphisms on a line yields the third. Together with identity
$\left(\mathrm{id}\right)$ this forms the group $\mathbb{Z}_{2}^{3}=\mathbb{Z}_{2}\times\mathbb{Z}_{2}\times\mathbb{Z}_{2}$
(right).}
\includegraphics[clip,scale=0.7]{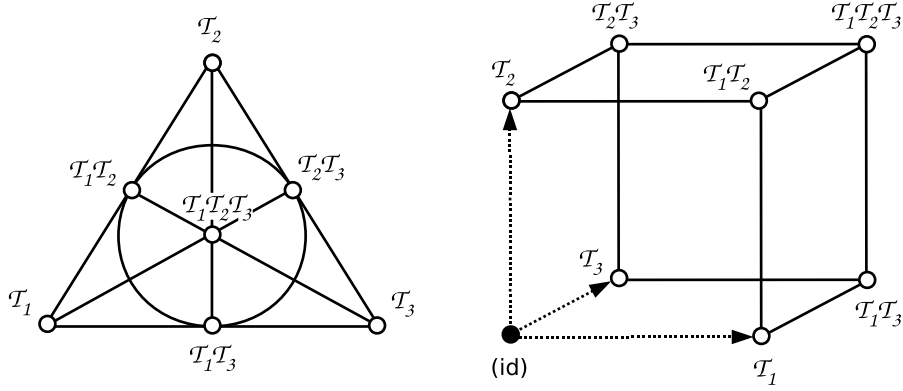}
\end{figure}

For a select $n$, the pair $\left\{ \left(\mathrm{id}\right),\mathcal{T}_{n}\right\} $
forms the two element cyclic group $\mathbb{Z}_{2}$ under repeat
application. The possible unique combinations of the $\left\{ \mathcal{T}_{1},\mathcal{T}_{2},\mathcal{T}_{3}\right\} $
form the set
\begin{equation}
\left\{ \mathcal{T}_{1},\,\mathcal{T}_{2},\,\mathcal{T}_{3},\,\mathcal{T}_{1}\mathcal{T}_{2},\,\mathcal{T}_{1}\mathcal{T}_{3},\,\mathcal{T}_{2}\mathcal{T}_{3},\,\mathcal{T}_{1}\mathcal{T}_{2}\mathcal{T}_{3}\right\} 
\end{equation}
which transitions between octonions $\mathbb{O}\left[N\right]$ of
the same chirality. It can be graphed in the Fano plane, where three
automorphisms lay on each line such that the combination of any two
automorphisms yields the third one (figure \ref{fig:T1T2T3Fano}).
 Together with the identity element, $\left(\mathrm{id}\right)$,
this forms the group $\mathbb{Z}_{2}^{3}=\mathbb{Z}_{2}\times\mathbb{Z}_{2}\times\mathbb{Z}_{2}$.

All possible combinations of the $\left\{ \mathcal{T}_{n}\right\} $
acting on $t_{\mathbb{O}\left[0\right]}$ then generate the 16 triplet
sets $t_{\mathbb{O}\left[N\right]}$ for the $\mathbb{O}\left[N\right]$
respectively. By choice, the chirality of all seven triplets in $t_{\mathbb{O}\left[0\right]}$
is chosen to be positive and written in abbreviated form as ``$+$''.
An explicit choice for the first eight $t_{\mathbb{O}\left[N\right]}$
is:
\begin{eqnarray}
t_{\mathbb{O}\left[0\right]} & := & \left\{ +,+,+,+,+,+,+\right\} =\left(\mathrm{id}\right)t_{\mathbb{O}\left[0\right]},\\
t_{\mathbb{O}\left[1\right]} & := & \left\{ +,+,-,-,+,-,-\right\} =\mathcal{T}_{3}t_{\mathbb{O}\left[0\right]},\nonumber \\
t_{\mathbb{O}\left[2\right]} & := & \left\{ +,-,+,-,-,+,-\right\} =\mathcal{T}_{2}t_{\mathbb{O}\left[0\right]},\nonumber \\
t_{\mathbb{O}\left[3\right]} & := & \left\{ +,-,-,+,-,-,+\right\} =\mathcal{T}_{2}\mathcal{T}_{3}t_{\mathbb{O}\left[0\right]},\nonumber \\
t_{\mathbb{O}\left[4\right]} & := & \left\{ -,-,-,-,+,+,+\right\} =\mathcal{T}_{1}t_{\mathbb{O}\left[0\right]},\nonumber \\
t_{\mathbb{O}\left[5\right]} & := & \left\{ -,-,+,+,+,-,-\right\} =\mathcal{T}_{1}\mathcal{T}_{3}t_{\mathbb{O}\left[0\right]},\nonumber \\
t_{\mathbb{O}\left[6\right]} & := & \left\{ -,+,-,+,-,+,-\right\} =\mathcal{T}_{1}\mathcal{T}_{2}t_{\mathbb{O}\left[0\right]},\nonumber \\
t_{\mathbb{O}\left[7\right]} & := & \left\{ -,+,+,-,-,-,+\right\} =\mathcal{T}_{1}\mathcal{T}_{2}\mathcal{T}_{3}t_{\mathbb{O}\left[0\right]}.\nonumber 
\end{eqnarray}
These correspond to the eight left-handed multiplication tables from
\cite{LockyerOctonions-Feb2008}. The right-handed ones are then obtained
from:
\begin{eqnarray}
t_{\mathbb{O}\left[N+8\right]} & := & \mathcal{T}_{0}t_{\mathbb{O}\left[N\right]}.
\end{eqnarray}

Overall, this concept is identical to the group action $T$ from \cite{SchrayManogueOcts1994}
(equation 30 therein). Octonions that are here mapped through $\mathcal{T}_{0}$
are called ``opposite algebra'' in \cite{SchrayManogueOcts1994}
(equation 33 therein), and correspond to octonionic spinors of opposite
chirality. The structure of octonion algebra and its relation to $\mathbb{Z}_{2}^{3}$
and Hadamard transforms is also investigated in \cite{AlbuMajidQuasialg}.

\section{Functions, distances, and ``octonion variance sieve''}

This section now defines a set of 16 polynomial functions, $A\left(f\right):=\left\{ f\left[N\right]\right\} $,
that use octonion multiplication rules $\mathbb{O}\left[N\right]$
on a given polynomial $f$. A certain linear superposition of these
functions, the Hadamard transform, will yield 16 ``distances'' $B\left(f\right)$
(corresponding to the 14 ``distances'' and two ``invariants''
from \cite{LockyerOctonions-Feb2008}). Applying the same superposition
rule again on the distances yields the original functions, making
them dual to each other. Furthermore, the automorphisms on the distances
$B\left(f\right)$ are similar to the automorphisms on the octonion
rules $\mathbb{O}\left[N\right]$ used to generate the functions $A\left(f\right)$.%
\footnote{On a sidenote, summing over results obtained from different multiplication
rules is also part of the two dimensional ``W space'' \cite{ShusterKoeplWSpace}.%
}

Let $f\in P$ be polynomial with finite number of arguments, $A$
the functor that turns the polynomial into a polynomial function in
$\mathbb{R}^{8}$, then $f\left[N\right]$ are the functions that
use the multiplication rule $\mathbb{O}\left[N\right]$ for multiplication:
\begin{eqnarray}
\left\{ f\left[N\right]\right\}  & := & \left\{ \mathbb{R}^{8}\otimes\ldots\otimes\mathbb{R}^{8}\rightarrow\mathbb{R}^{8},\, f\left[N\right]\in\mathbb{O}\left[N\right]\right\} ,\\
A & : & P\rightarrow\left\{ \mathbb{R}^{8}\otimes\ldots\otimes\mathbb{R}^{8}\rightarrow\mathbb{R}^{8}\right\} .\nonumber 
\end{eqnarray}
An $f\left[N\right]$ can then be represented by the resultant vector
made from general coefficients of the function's parameters. Because
octonions are a normed division algebra, with unique multiplicative
inverses and free from zero-divisors or nilpotents (except $0$),
knowledge of all coefficients from a general octonion product allows
to uniquely identify the multiplication rule $t_{\mathbb{O}\left[N\right]}$
used%
\footnote{Except of course for the trivial case where no octonion multiplication
occurred at all, such as e.g. $f\left(a_{0},a_{1}\right)=a_{0}+a_{1}$.%
}.

Given a polynomial $f$, all functions $f\left[N\right]$ form the
set $A\left(f\right)$:
\begin{eqnarray}
A\left(f\right) & := & \left\{ f\left[N\right]\right\} ,\qquad N=0,\ldots,15.
\end{eqnarray}
The automorphisms $S^{A}$ on $A\left(f\right)$ follow directly from
the construction of the $t_{\mathbb{O}\left[N\right]}$ above, as
the group of repeat application of the $\mathcal{T}_{n}$ and identity
$\left(\mathrm{id}\right)$ on the associative triplets $t_{\mathbb{O}\left[N\right]}$:
\begin{eqnarray}
S^{A} & : & A\left(f\right)\rightarrow A\left(f\right),\nonumber \\
S^{A} & \cong & \left\{ \left(\mathrm{id}\right),\mathcal{T}_{0}\right\} \times\left\{ \left(\mathrm{id}\right),\mathcal{T}_{1}\right\} \times\left\{ \left(\mathrm{id}\right),\mathcal{T}_{2}\right\} \times\left\{ \left(\mathrm{id}\right),\mathcal{T}_{3}\right\} =\mathbb{Z}_{2}^{4}.\label{eq:automorphismsOnA}
\end{eqnarray}

Distances $B\left(f\right)$ are now constructed from linear superposition
of the 16 functions in $A\left(f\right)$. Left- and right-octonions
(with $N<8$ and $N\geq8$ respectively) will not be treated separately
as in \cite{LockyerOctonions-Feb2008}, instead they will be handled
as one set here. A sign matrix is defined:
\begin{eqnarray}
b_{jk} & := & \left(-1\right)^{j\wedge k},\qquad j,k=0,\ldots,15,
\end{eqnarray}
where $j\wedge k$ is logical ``and'' from bitwise representation
of the $j$ and $k$. The distances $B\left(f\right)$ then are the
Hadamard transforms $H_{4}$ on the functions:
\begin{eqnarray}
g\left[k\right] & := & \frac{1}{4}\sum_{j=0}^{15}b_{jk}f\left[j\right],\qquad k=0,\ldots,15,\label{eq:transformFunctions}\\
B\left(f\right) & := & \left\{ g\left[k\right]\right\} =\left\{ \frac{1}{4}\sum_{j=0}^{15}b_{jk}f\left[j\right]\right\} =H_{4}\left(f\left[N\right]\right).\nonumber 
\end{eqnarray}
Since
\begin{eqnarray}
\frac{1}{4}\sum_{k=0}^{15}b_{kl}g\left[k\right] & = & \frac{1}{16}\sum_{j,k=0}^{15}b_{jk}b_{kl}f\left[j\right]=\frac{1}{16}\sum_{j,k=0}^{15}\left(-1\right)^{j\wedge k}\left(-1\right)^{k\wedge l}f\left[j\right]\label{eq:transformDistances}\\
 & = & \frac{1}{16}\sum_{j,k=0}^{15}\left(-1\right)^{k\wedge\left(j+l\right)}f\left[j\right]=\frac{1}{16}\sum_{j,k=0}^{15}b_{k\left(j+l\right)}f\left[j\right]=f\left[l\right].\nonumber 
\end{eqnarray}
the sets of distances $B\left(f\right)$ and functions $A\left(f\right)$
are dual to each other, related through linear superposition using
the sign matrix $b_{jk}$. This dualism is a property of the Hadamard
transform in general, as it is its own inverse.

Rows and columns in the $b_{jk}$ correspond to functions $f\left[k\right]$
and distances $g\left[j\right]$. The set of rows $\left\{ b_{jk},\, k\textrm{ fixed}\right\} $
can be constructed from a 16 element vector $b_{j0}$ and four duality
morphisms $\left\{ \mathcal{T}_{n}^{b}\right\} $ acting on the sign
of $b_{j0}$: 
\begin{eqnarray}
b_{j0} & = & \left(1,1,1,1,1,1,1,1,1,1,1,1,1,1,1,1\right),\\
\mathcal{T}_{0}^{b} & := & \left(+,+,+,+,+,+,+,+,-,-,-,-,-,-,-,-\right),\nonumber \\
\mathcal{T}_{1}^{b} & := & \left(+,+,+,+,-,-,-,-,+,+,+,+,-,-,-,-\right),\nonumber \\
\mathcal{T}_{2}^{b} & := & \left(+,+,-,-,+,+,-,-,+,+,-,-,+,+,-,-\right),\nonumber \\
\mathcal{T}_{3}^{b} & := & \left(+,-,+,-,+,-,+,-,+,-,+,-,+,-,+,-\right).\nonumber 
\end{eqnarray}
All possible combinations of the $\left\{ \mathcal{T}_{n}^{b}\right\} $
and identity $\left(\mathrm{id}\right)$ on $b_{j0}$ generate the
set of rows $\left\{ b_{jk},\, k\textrm{ fixed}\right\} $ (and since
$b_{jk}=b_{kj}$ the similar reasoning applies to the columns $\left\{ b_{jk},\, j\textrm{ fixed}\right\} $).
The actual order of rows (or columns) is not significant. Simultaneously
swapping any two rows and columns with same indices always yields
another Hadamard transform.

Given a set of functions $\left\{ f\left[N\right]\right\} $, the
automorphisms $S^{B}$ on the distances $B\left(f\right)$ then are:
\begin{eqnarray}
S^{B} & : & B\left(f\right)\rightarrow B\left(f\right),\nonumber \\
S^{B} & \cong & \left\{ \left(\mathrm{id}\right),\mathcal{T}_{0}^{b}\right\} \times\left\{ \left(\mathrm{id}\right),\mathcal{T}_{1}^{b}\right\} \times\left\{ \left(\mathrm{id}\right),\mathcal{T}_{2}^{b}\right\} \times\left\{ \left(\mathrm{id}\right),\mathcal{T}_{3}^{b}\right\} =\mathbb{Z}_{2}^{4}.\label{eq:automorphismsOnB}
\end{eqnarray}
This makes the automorphisms $S^{A}$ on the octonion functions $A\left(f\right)$
similar to the automorphisms on the distances $B\left(f\right)$,
\begin{eqnarray}
S^{A} & \sim & S^{B}.
\end{eqnarray}

With the definitions from above, the ``octonion variance sieve''
from \cite{LockyerOctonions-Feb2008} is equivalent to computing the
$g\left[j\right]$. An ``algebraic invariant'' is then built from
functions $f\left[0\right]$ such that:
\begin{eqnarray}
g\left[j\right] & = & 0\textrm{ for }j>0.\label{eq:defAlgebraicInvariant}
\end{eqnarray}

\section{Algebra of Derivations}

The algebra of derivations $\mathfrak{der}\left(\mathbb{A}\right)$
in $\mathbb{A}\in\left\{ \mathbb{H},\mathbb{O}\right\} $ is the Lie
algebra that corresponds to the automorphism Lie group over $\mathbb{A}$:
\begin{eqnarray}
\mathfrak{der}\left(\mathbb{A}\right) & = & \begin{cases}
\mathfrak{so}\left(3\right) & \left(\mathbb{A}=\mathbb{H},\,\mathrm{Aut}\left(\mathbb{H}\right)\cong\mathrm{SO}\left(3\right)\right),\\
\mathfrak{g}_{2} & \left(\mathbb{A}=\mathbb{O},\,\mathrm{Aut}\left(\mathbb{O}\right)\cong\mathrm{G}_{2}\right).
\end{cases}
\end{eqnarray}
The $\mathfrak{der}\left(\mathbb{A}\right)$ have features similar
to a derivative operator:
\begin{eqnarray}
\mathfrak{der}\left(\mathbb{A}\right) & := & \left\{ D\,:\,\mathbb{A}\rightarrow\mathbb{A}\textrm{, where }D\left(ab\right)=D\left(a\right)b+aD\left(b\right),\, a,b\in\mathbb{A}\right\} .
\end{eqnarray}
Any $u,v\in\mathbb{A}$ form a $\mathfrak{der}\left(\mathbb{A}\right)$
as \cite{Baez2002TheOctonions,Schafer1995nonassIntro}:
\begin{eqnarray}
D_{u,v}\left(a\right) & := & \left[\left[u,v\right],a\right]-3\left(\left(uv\right)a-u\left(va\right)\right),\\
a,u,v & \in & \mathbb{A},\qquad\mathbb{A}\in\left\{ \mathbb{H},\mathbb{O}\right\} .\nonumber 
\end{eqnarray}

It is now shown that changing between equivalent octonion algebras
from above leaves such algebra of derivations invariant exactly in
their associative subspaces. Since $D_{u,v}\left(a\right)$ is linear
in its arguments $\left\{ a,u,v\right\} $ it is sufficient to examine
relations where $a,u,v$ are basis elements from $b_{\mathbb{O}}:=\left\{ 1,i_{1},\ldots,i_{7}\right\} $.
For real $u$, $v$ or $a$ the commutator brackets $\left[\left[u,v\right],a\right]$
and associator $\left(\left(uv\right)a-u\left(va\right)\right)$ are
always zero. This becomes the trivial case. If the $\left\{ a,u,v\right\} $
are part of an associative triplet then the associator is zero. This
is always the case for quaternions, where changing the order of all
multiplications does not change the remaining commutator brackets:
\begin{eqnarray}
\left[\left[u,v\right],a\right] & = & uva-vua-auv+avu\\
 & = & avu-auv-vua+uva\nonumber \\
 &  & \textrm{for }\left\{ a,u,v\right\} \in\mathbb{H}.\nonumber 
\end{eqnarray}
All quaternion algebras therefore have the same algebra of derivations
$\mathfrak{der}\left(\mathbb{H}\right)=\mathfrak{so}\left(3\right)$.

In the remaining case the $\left\{ a,u,v\right\} $ form an antiassociative
triplet. The three elements are pairwise anticommutative and each
element is also anticommutative with the product of the other two.
This yields:
\begin{eqnarray}
 &  & \left[\left[u,v\right],a\right]-3\left(\left(uv\right)a-u\left(va\right)\right)\label{eq:derivAlgebraAntiassociative}\\
 & = & \left(uv\right)a-\left(vu\right)a-a\left(uv\right)+a\left(vu\right)-3\left(uv\right)a+3u\left(va\right)\nonumber \\
 & = & \left(uv\right)a+\left(uv\right)a+\left(uv\right)a+\left(uv\right)a-3\left(uv\right)a-3\left(uv\right)a\nonumber \\
 & = & -2\left(uv\right)a.\nonumber 
\end{eqnarray}

By inspecting the algebra automorphisms $\mathcal{T}_{n}$ on the
set of equivalent octonion algebras (\ref{eq:defAutomorphismsT})
there are no two basis element triplets that simultaneously retain
or change parity across all four $\mathcal{T}_{n}$. This means that
there are no two triplets that contain $\left\{ u,v\right\} $ and
$\left\{ uv,a\right\} $, respectively, such that the product $\left(uv\right)a$
would remain unaltered in all 16 algebras mapped by the $\mathcal{T}_{n}$.
Therefore, the $D_{u,v}\left(a\right)$ can only yield the same algebra
for all 16 equivalent octonion multiplications $\mathbb{O}\left[N\right]$
in its associative subspace $\mathfrak{so}\left(3\right)\subset\mathfrak{g}_{2}$:
\begin{eqnarray}
D_{u,v}\left(a\right)\left[N\right]=D_{u,v}\left(a\right)\left[0\right]\textrm{ for }N=0,\ldots,15 & \, & \,
\end{eqnarray}
\begin{eqnarray*}
\, & \Longleftrightarrow & D_{u,v}\left(a\right)\left[0\right]\in\mathfrak{der}\left(\mathbb{H}\right).
\end{eqnarray*}

This is also the case if one generalizes $a\in\mathbb{O}$ to polynomial
functions $A\left(f\right)=\left\{ f\left[N\right]\right\} $. The
$-2\left(uv\right)a$ term from equation (\ref{eq:derivAlgebraAntiassociative})
cannot be made to vanish from any term in a polynomial function $f\left[N\right]$
that doesn't associate with $u$ and $v$ since octonions are free
from zero-divisors and nilpotents. The $f\left[N\right]$ therefore
must be contained in the subalgebra of quaternionic polynomials $h:\mathbb{H}\rightarrow\mathbb{H}$
which include $u$ and $v$ as well, to have the same derivation algebra
for any $N$:
\begin{eqnarray}
D_{u,v}\left(f\right)\left[N\right]=D_{u,v}\left(f\right)\left[0\right]\textrm{ for }N=0,\ldots,15 & \, & \,
\end{eqnarray}
\begin{eqnarray*}
\, & \Longleftrightarrow & \left\{ f\left[N\right]\right\} \in\left\{ h:\mathbb{H}_{u,v}\rightarrow\mathbb{H}_{u,v}\right\} ;\, u,v\in\mathbb{H}_{u,v}.
\end{eqnarray*}
With this, ``algebraic invariance'' condition from equation (\ref{eq:defAlgebraicInvariant})
can be written as:
\begin{eqnarray}
\mathfrak{der}\left(f\right) & \overset{!}{\subseteq} & \mathfrak{der}\left(\mathbb{H}\right).
\end{eqnarray}

For octonionic differential expressions $\hat{D}f$ an operator $\hat{D}$
may exist such that: 
\begin{eqnarray}
\mathfrak{der}\left(\hat{D}f\right) & \subseteq & \mathfrak{der}\left(\mathbb{H}\right).\label{eq:notNecessarilyQuaternionic}
\end{eqnarray}
If the ``octonion variance sieve process'' is indeed as claimed
\cite{LockyerOctonions-Feb2008}, then this does not necessarily require
$f$ to be quaternionic.
\begin{rem}
\label{remarkExoticR4}If true, this construction could relate to
exotic $\mathbb{R}^{4}$ spaces. Solutions for equation (\ref{eq:notNecessarilyQuaternionic})
are generally quaternionic, which are built over a set that is homeomorphic
to $\mathbb{R}^{4}$. However, due to nonassociativity of octonion
multiplication $\hat{D}f$, the various spaces $\mathfrak{der}\left(\hat{D}f\right)$
might not necessarily be diffeomorphic to $\mathfrak{der}\left(\mathbb{H}\right)$.
The construction that led to equation (\ref{eq:notNecessarilyQuaternionic})
requires nonassociativity of a normed division algebra, uniquely satisfied
by octonions, which makes its solutions specific to spaces over $\mathbb{R}^{4}$.
More investigation is needed to better understand the claim and its
consequences, if true.
\end{rem}

\section*{Acknowledgments}

\thanks{Many thanks to Armahedi Mahzar for valuable discussion, demonstrating
the $\mathbb{Z}_{2}^{3}$ structure of figure \ref{fig:T1T2T3Fano}
with its relation to the Klein four-group $\mathbb{Z}_{2}^{2}$, and
that set of automorphisms on the $\left\{ \mathbb{O}\left[N\right]\right\} $
is forming the reflection group in four dimensions, $C_{2}^{4}$,
the ``Klein 16-group''. Also, many thanks to Rick Lockyer for explaining
his work in detail and demonstrating that the $b_{jk}$ generate a
Hadamard transform. Furthermore, thanks to Joy Christian for keeping
me honest about correcting the 2011 version of this paper, and to
Ben Dribus for speculating that some of my work may be related to
exotic $\mathbb{R}^{4}$ spaces.}

\section*{Glossary of symbols}
\begin{lyxlist}{00.00.0000}
\item [{$b_{\mathbb{O}}$}] An ordered set of orthogonal vectors $\left\{ 1,i_{1},\ldots,i_{7}\right\} \in\mathbb{R}^{8}$.
\item [{$\mathbb{O}\left[N\right]$}] An octonion algebra with a given
a $b_{\mathbb{O}}$ as basis and one of 16 multiplication rules indexed
with $N=0,\ldots,15$.
\item [{$t_{\mathbb{O}\left[N\right]}$}] The seven associative triplets
of basis elements from $\mathbb{O}\left[N\right]$.
\item [{$f\left[N\right]$}] Polynomial functions on $\mathbb{R}^{8}$,
$f\left[N\right]:\mathbb{R}^{8}\times\ldots\times\mathbb{R}^{8}\rightarrow\mathbb{R}^{8}$,
that use $\mathbb{O}\left[N\right]$ for multiplication.
\item [{$A\left(f\right)$}] Given a polynomial $f$, the set of functions
$A\left(f\right):=\left\{ f\left[N\right]\right\} ,\, N=0,\ldots,15$.
\item [{$\mathcal{T}_{0},\ldots,\mathcal{T}_{3}$}] Duality automorphisms
on the multiplication rules, $\mathcal{T}_{n}:\left\{ \mathbb{O}\left[N\right]\right\} \rightarrow\left\{ \mathbb{O}\left[N\right]\right\} $,
$\mathcal{T}_{n}\mathcal{T}_{n}=\left(\mathrm{id}\right)$ (for $n=0,1,2,3$).
\item [{$S^{A}$}] The automorphisms on $A\left(f\right)$, i.e.: $S^{A}:A\left(f\right)\rightarrow A\left(f\right)$.
\item [{$b_{jk}$}] A $16\times16$ sign matrix, $b_{jk}:=\left(-1\right)^{j\wedge k}$;
$j,k=0,\ldots,15$.
\item [{$H_{4}$}] Hadamard transform generated by the $b_{jk}$.
\item [{$g\left[N\right]$}] A ``distance'' function obtained from linear
superposition of the $f\left[N\right]$, using $b_{jk}$ as coefficients.
\item [{$B\left(f\right)$}] Given a polynomial $f$, the set of distance
functions $B\left(f\right):=H_{4}\left(A\left(f\right)\right)=\left\{ g\left[N\right]\right\} ,\, N=0,\ldots,15$.
\item [{$\mathcal{T}_{0}^{b},\ldots,\mathcal{T}_{3}^{b}$}] Duality automorphisms
on the rows (or columns) of $b_{jk}$.
\item [{$S^{B}$}] The automorphisms on $B\left(f\right)$, i.e.: $S^{B}:B\left(f\right)\rightarrow B\left(f\right)$.
\item [{$\mathfrak{der}\left(\mathbb{A}\right)$}] Algebra of derivations
over $\mathbb{A}$, e.g. $\mathfrak{der}\left(\mathbb{H}\right)=\mathfrak{so}\left(3\right)$,
$\mathfrak{der}\left(\mathbb{O}\right)=\mathfrak{g}_{2}$.\end{lyxlist}

\end{document}